\documentclass[aip,apl,reprint]{revtex4-1}

\usepackage{graphicx}

\begin{document}

\title{Growth of atomically smooth thin films of the electronically phase separated manganite (La$_{0.5}$Pr$_{0.5}$)$_{0.67}$Ca$_{0.33}$MnO$_{3}$} 


\author{Hyoungjeen Jeen}
\altaffiliation[Presently at: ]{Materials Science and Technology Division, Oak Ridge National Laboratory, Oak Ridge, Tennessee 37831, USA}
\author{Rafiya Javed}
\author{Amlan Biswas}
\email{amlan@phys.ufl.edu}
\affiliation{Department of Physics, University of Florida, Gainesville, Florida 32611, USA}


\date{\today}

\begin{abstract}
Atomically flat, epitaxial, and stoichiometric thin films of the electronically phase separated compound (La$_{0.5}$Pr$_{0.5}$)$_{0.67}$Ca$_{0.33}$MnO$_{3}$  were grown on as-received and treated NdGaO$_{3}$ substrates by fine tuning of oxygen pressure during deposition. Optimal thin films with step flow growth mode show superior physical properties compared to thin films grown in off-optimal oxygen pressures, {\em viz.} the highest maximum temperature coefficient of resistance, the highest peak-resistivity temperature, and reduced coercive fields. Transport, magnetization, and x-ray diffraction measurements indicate that the oxygen pressure during growth plays a critical role in the formation of oxygen vacancies, cation vacancies, and grain boundaries.     
\end{abstract}

\pacs{}

\maketitle 
   Manganites show colossal responses to external fields, since their structure, transport, and magnetism are closely related ~\cite{jin94,ueh99,ahn04}. When grown in the thin film form, manganites also exhibit unique properties associated with substrate induced strain ~\cite{Bis00,Suz98,War09}. One aim in thin film research is to achieve physical properties similar to bulk materials using methods such as post annealing in an oxygen rich environment and growing films in highly energetic, reducing conditions~\cite{Cho09,Pre01, mar10}. However, the discovery of physical properties unique to thin films and not found in bulk forms of the same compounds has also been a driver for thin film research. Examples of such properties are anisotropic transport, substrate induced metal-insulator transition, in-plane magnetic anisotropy, and insulator-metal transition in manganite superlattices ~\cite{War09, Bis00, Bos09, Bha08}. The interface region between two perovskite materials is also a subject of recent interest since this region has properties not observed in the bulk form of either of the constituent materials ~\cite{Sma07, Oht04}. Observation and identification of such properties requires careful optimization of thin film growth conditions. For interfaces between two insulating perovskites such optimization involves mainly the structural properties~\cite{Ari11}. If we want to study the properties of magnetic and/or metallic thin films and interfaces, the transport and magnetism need to be optimized along with the structure. Thus, the fabrication of atomically flat, epitaxial, and stoichiometric manganite thin films is an essential step toward the search for unique physical properties arising due to the effect of strain on the magnetism and phase competition at thin film interfaces.    
   
   Thin film growth of materials such as (La$_{1-y}$Pr$_{y}$)$_{1-x}$Ca$_{x}$MnO$_{3}$ is complicated by the structural phase separation at low temperatures~\cite{ueh99}. Thus, thin films which show excellent structural properties at room temperature may not show the expected transport and magnetic properties at low temperature due to the presence of the substrate~\cite{War08, War09}. We show that atomically flat, epitaxial, and stoichiometric (La$_{0.5}$Pr$_{0.5}$)$_{0.67}$Ca$_{0.33}$MnO$_{3}$ (LPCMO) thin films on (110) NdGaO$_{3}$ (NGO) substrates can be grown with optimized magnetic and transport properties using fine control of oxygen pressure and without post-annealing. This optimization was confirmed by various physical property measurements, \textit{viz.} surface morphology, transport, structural, and magnetic measurements. The LPCMO thin films, grown under optimal oxygen pressure, show atomic steps, the highest value for the maximum temperature coefficient of resistivity ($TCR_{max}$), and the highest peak resistivity temperature ($T_{P}$). These films also show reduced coercive fields at low temperatures with magnetization vs. field ($M-H$) loops typical of materials with in-plane magnetic anisotropy~\cite{jee11}. We propose that oxygen pressure affects the valency of the Mn ions due to oxygen vacancies, cation vacancies, and creation/reduction of grain boundaries~\cite{ju95,Pre99,Pre01}.      

\begin{figure}[b]
\includegraphics[width=8.5cm] {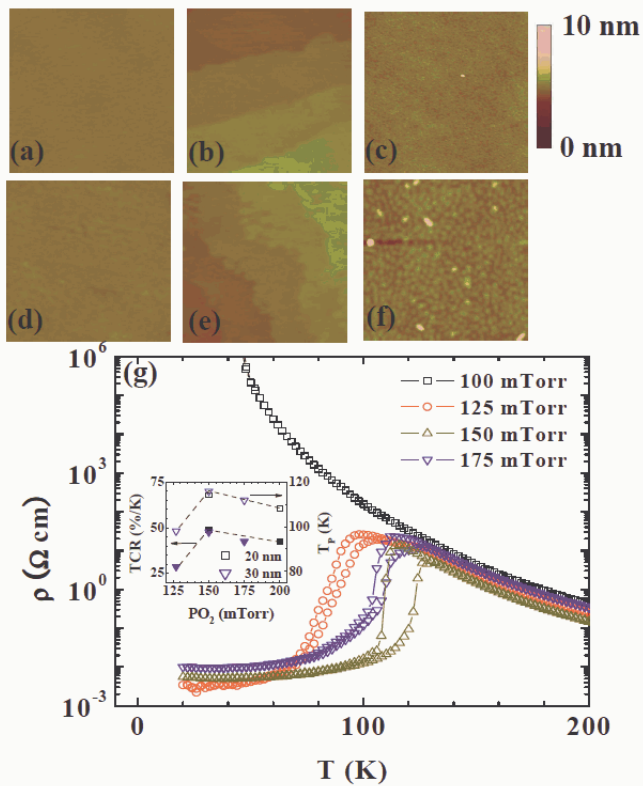}%
\caption{(Color online) 1 $\mu$m $\times$ 1$\mu$m AFM images of (a) as-received NGO substrate and (b) annealed NGO substrate. AFM images of LPCMO thin films grown on as-received NGO in oxygen pressures of (c) 100 mTorr, (d) 125 mTorr, (e) 150 mTorr, and (f) 175 mTorr. (g) $\rho(T)$ of the LPCMO thin films grown in different oxygen pressures. Inset: $TCR_{max}$ and $T_{P}$ as a function of oxygen pressure (PO$_{2}$). All dotted lines are guides for eyes. }%
\end{figure}
LPCMO thin films were grown on two different substrates, \textit{viz.} as-received epi-polished (110) NGO substrates and subsequently thermally treated (110) NGO substrates. The thermal treatment was carried out in air to create atomically flat surface and steps~\cite{lec03}. The substrates were first heated at 20$^{\circ}$C/min to 600$^{\circ}$C, then at 10$^{\circ}$C/min to 950$^{\circ}$C. After heat treatment at 950$^{\circ}$C for 30 minutes, the substrates were then cooled to 550$^{\circ}$C at 12$^{\circ}$C/min and then the furnace was switched off. Atomic force microscope (AFM) images of the treated substrates consistently showed atomic steps (about 0.4 nm) and low miscut angles (about 0.05$^{\circ}$) (Fig. 1 (a) and (b)). We deposited LPCMO thin films on the as-received and treated substrates using pulsed laser deposition using the conditions described in Jeen \textit{et al.}.~\cite{jee11}. However, the oxygen pressure was varied by design during each deposition from 75 mTorr to 200 mTorr. The films were grown to a thickness of 30 nm. After deposition, the film was cooled down at 20$^{\circ}$C/min in an oxygen pressure of 440 Torr. Transport measurements were performed using DC polarity reversal method~\cite{kei07} with four probe wiring and low current (5 nA) to minimize current/voltage effect on the manganite thin films ~\cite{dha07}. Magnetization measurements~\cite{jee11} and $\theta-2\theta$ x-ray diffraction (XRD) were performed using a Quantum design 5 T SQUID magnetometer and Philips APD 3720, respectively.

     We observed differences in surface morphology in LPCMO thin films grown under different oxygen pressures (PO$_2$) (Fig. 1 (c)-(f)). Smooth surfaces with r.m.s. roughness of 0.5 nm were observed in films grown in PO$_2$ $\leq$ 100 mTorr. However, we could not find any characteristic features such as islands or steps. On increasing PO$_2$ to 150 mTorr, atomic steps were formed due to step flow growth mode. The atomic step height was about 0.4 nm which is close to the size of an LPCMO unit cell~\cite{Col03}. For PO$_2$ $\geq$ 175 mTorr, the thin films followed a 2D island growth mode. The r.m.s. roughness was still less than 0.5 nm. These films are different from LPCMO thin films showing 3D island growth mode, which were grown with high laser fluence (1.0 $\pm$ 0.2 J/cm$^2$) and PO$_2$ (420 mTorr) but showed desirable transport properties~\cite{tos09}. In the oxygen pressure range we investigated, interlayer mass transport of LPCMO was well executed, since the r.m.s. roughness of all samples was close to the unit cell size of LPCMO ~\cite{Chr08}. The films grown in optimal PO$_2$ (150 mTorr) show atomic steps and the highest $TCR_{max}$ and $T_{P}$ (inset of Fig. 1 (g)). Films grown in PO$_2$ $<$ 100 mTorr show insulating behavior down to the lowest measurement temperature. These trends suggest a change of the Mn$^{4+}$/Mn$^{3+}$ ratio~\cite{ju95} since the non-monotonic behavior of $TCR_{max}$ and $T_{P}$ as a function of oxygen pressure can be explained by a change in the valency of Mn-ion~\cite{Uru95,Sch95}. For our films grown in less than 150 mTorr, the average valency of the Mn-ions could be less than the expected value of +3.67 due to oxygen vacancies. For PO$_2$ $>$ 150 mTorr, the average valency of the Mn-ions is greater than +3.67 due to cation vacancies rather than oxygen abundance, since oxygen rich manganites are hard to grow ~\cite{ ju95, Pre99, mar10, Org10}. A change in the ratio between La and Pr ion concentrations is unlikely since such a change would lead to a monotonic shift of $T_{P}$ rather than observed shape. Grain boundary effects may also contribute to the low $TCR_{max}$, low $T_{P}$, and higher resistivity at low temperatures of the films grown in oxygen rich condition, since such films show granular morphology (Fig. 1 (f))~\cite{Hwa96}.     
\begin{figure}[b]
\includegraphics[width=8.5cm] {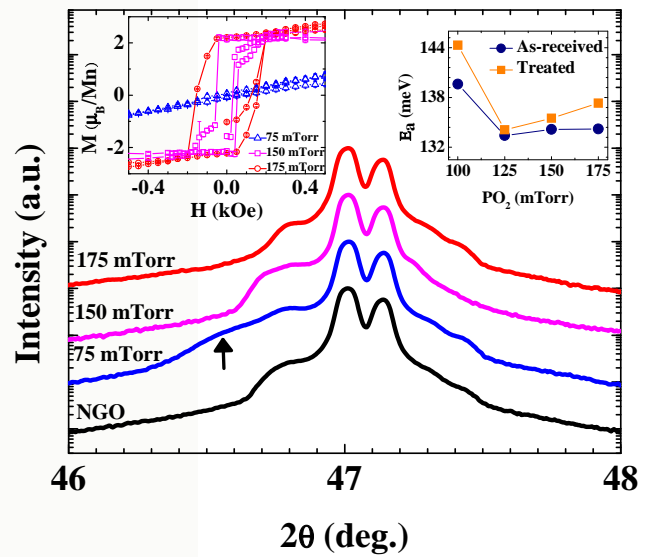}%
\caption{(Color online) $\theta$-2$\theta$ XRD patterns of the LPCMO thin films grown in different oxygen pressures. The curves have been shifted vertically for clarity. The black arrow indicates the diffuse LPCMO peak for the film grown in 75 mTorr oxygen. Left inset: Magnetic moment as a function of magnetic field, along [1$\bar{1}$0] (easy axis) direction of LPCMO thin films on as-received NGO. Right inset: Activation energies of LPCMO films grown on as-received NGO and treated NGO.}%
\end{figure}

  Next we measured the magnetization and structural properties of the films. An increase of the lattice constant and reduction in magnetization are typically observed in oxygen deficient but optimally doped manganites ~\cite{ju95, Pre99}. We observed diffuse and broad XRD peaks in the film grown in 75 mTorr O$_{2}$ pressure (Fig. 2), while the XRD patterns of LPCMO thin films with either atomic steps or 2D islands overlapped with the pattern of the nearly lattice-matched NGO substrate. We also observed a pronounced reduction of remanent magnetization and disappearance of magnetic hysteresis in the oxygen deficient film (Left inset of Fig. 2). The films with atomic steps have lower coercive fields and sharper changes of magnetic moments near the coercive fields than the film with 2D islands. These differences in coercive fields may be due to grain boundary effects in the film with 2D islands~\cite{Tan02}. Since the XRD patterns for all the films showed only minor differences, we can conclude that while XRD measurements can be used for determining conditions which are far from optimal, it cannot be used for accurately determining the optimal conditions. 

In an attempt to trim the optimal conditions further, we analyzed the transport properties of the LPCMO thin films in the high temperature region. We calculated the activation energies ($E_a$) of the films from transport data above $T_P$. The adiabatic small polaron model and the variable range hopping model were employed to fit the high temperature data with the small polaron model providing the better fit~\cite{Sal01,Zie98a}. However, the $E_a$ values do not show a significant variation among the thin films grown in different oxygen pressures (Right inset of Fig. 2). $E_a$ $\approx$ 133.5 $\pm$ 0.5 meV for the films which show a metal-insulator transition with either step flow growth mode or island growth mode. The films grown in PO$_2$ = 100 mTorr do not show an insulator to metal (I-M) transition and have a higher $E_a$ ($\approx$144 meV). Oxygen deficient manganite films have an average Mn-ion valency less than +3.67 due to oxygen vacancies, which could hamper the double exchange mechanism thus removing the I-M transition and increasing $E_a$ ~\cite{ju95}. From the measurements and analysis above, it is noted that the LPCMO thin films with atomic steps grown in optimal oxygen pressure of 150 mTorr can be distinguished from the films grown in off-optimal oxygen pressures using low temperature transport and magnetization measurements, since only the films with atomic steps show both lower coercive fields and a sharp insulator to metal transition.~\cite{Tan02}.
\begin{figure}[t]
\includegraphics[width=8.5cm] {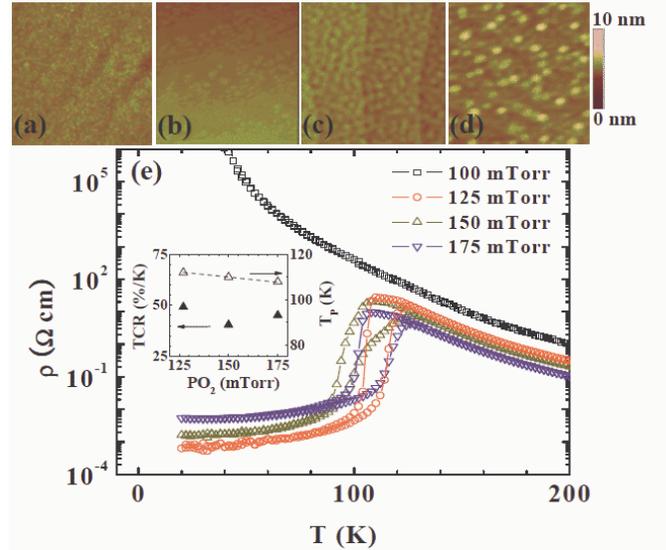}%
\caption{(Color online) 1 $\mu$m $\times$ 1 $\mu$m AFM images of LPCMO thin films on treated NGO grown in oxygen pressures of (a) 100 mTorr, (b) 125 mTorr, (c) 150 mTorr, and (d) 175 mTorr. (e) $\rho(T)$ of the LPCMO thin films. Inset: $TCR_{max}$ and $T_{P}$ as a function of PO$_{2}$. The dotted line is a guide for the eye.}%
\end{figure}     
  
    We also confirmed similar trends in surface morphology, transport, XRD, and magnetization measurements in LPCMO thin films grown on treated NGO (tNGO) substrates. The atomic steps in the LPCMO thin films followed the atomic steps on the tNGO substrates (Fig. 1 (b) and figure 3 (b)), which indicates that the step edges on tNGO provide energetically favorable sites for thin film growth~\cite{Chr08}. The trends of $TCR_{max}$ and $T_{P}$ are similar in LPCMO thin films on treated and as-received NGO (Inset of Fig. 3 (e)). However, the optimal PO$_2$ for step flow growth mode is shifted down to 125 mTorr for the tNGO substrates. The $E_a$ of the thin films showing I-M transition is about 135 $\pm$ 1 meV and for the insulating films grown in PO$_2$ = 100 mTorr, $E_a$ $\approx$ 140 meV (Right inset of Fig. 2). We also observed a diffuse XRD peak at a lower angle and lower magnetization in insulating LPCMO thin films grown in PO$_2$ = 75 mTorr  similar to the observations for the films grown on as-received substrates. 
When the films grown on the two types of substrates are compared, the optimized films grown on tNGO show no resistivity upturn at low temperature unlike the optimized films grown on the as-received substrates. Since substrate induced strain favors the metallic phase in LPCMO thin films grown on NGO ~\cite{dha07}, the absence of the low temperature resistivity upturn suggests that the LPCMO thin films grown on tNGO conform better to the substrates and have lower disorder~\cite{dha07,yun08}. 

	  In summary, we have successfully grown atomically flat, epitaxial, and stoichiometric LPCMO thin films on NGO substrates by tuning the oxygen pressure during growth. Surface morphology, transport and magnetization measurements are the critical measurements for determining the optimum film growth conditions~\cite{Org10} since, while necessary, activation energies and XRD measurements are not enough to distinguish the film properties. Disorder in thin films is reduced when treated substrates with singly terminated, atomically smooth surfaces are used. 

	This work was supported by NSF DMR-0804452. The authors are grateful for discussion with H. N. Lee. X-ray diffraction was conducted at major analytical instrumentation center, University of Florida. 


\begin{thebibliography}{32}%
\makeatletter
\providecommand \@ifxundefined [1]{%
 \@ifx{#1\undefined}
}%
\providecommand \@ifnum [1]{%
 \ifnum #1\expandafter \@firstoftwo
 \else \expandafter \@secondoftwo
 \fi
}%
\providecommand \@ifx [1]{%
 \ifx #1\expandafter \@firstoftwo
 \else \expandafter \@secondoftwo
 \fi
}%
\providecommand \natexlab [1]{#1}%
\providecommand \enquote  [1]{``#1''}%
\providecommand \bibnamefont  [1]{#1}%
\providecommand \bibfnamefont [1]{#1}%
\providecommand \citenamefont [1]{#1}%
\providecommand \href@noop [0]{\@secondoftwo}%
\providecommand \href [0]{\begingroup \@sanitize@url \@href}%
\providecommand \@href[1]{\@@startlink{#1}\@@href}%
\providecommand \@@href[1]{\endgroup#1\@@endlink}%
\providecommand \@sanitize@url [0]{\catcode `\\12\catcode `\$12\catcode
  `\&12\catcode `\#12\catcode `\^12\catcode `\_12\catcode `\%12\relax}%
\providecommand \@@startlink[1]{}%
\providecommand \@@endlink[0]{}%
\providecommand \url  [0]{\begingroup\@sanitize@url \@url }%
\providecommand \@url [1]{\endgroup\@href {#1}{\urlprefix }}%
\providecommand \urlprefix  [0]{URL }%
\providecommand \Eprint [0]{\href }%
\providecommand \doibase [0]{http://dx.doi.org/}%
\providecommand \selectlanguage [0]{\@gobble}%
\providecommand \bibinfo  [0]{\@secondoftwo}%
\providecommand \bibfield  [0]{\@secondoftwo}%
\providecommand \translation [1]{[#1]}%
\providecommand \BibitemOpen [0]{}%
\providecommand \bibitemStop [0]{}%
\providecommand \bibitemNoStop [0]{.\EOS\space}%
\providecommand \EOS [0]{\spacefactor3000\relax}%
\providecommand \BibitemShut  [1]{\csname bibitem#1\endcsname}%
\let\auto@bib@innerbib\@empty
\bibitem [{\citenamefont {Jin}\ \emph {et~al.}(1994)\citenamefont {Jin},
  \citenamefont {Tiefel}, \citenamefont {McCormack}, \citenamefont {Fastnacht},
  \citenamefont {Ramesh},\ and\ \citenamefont {Chen}}]{jin94}%
  \BibitemOpen
  \bibfield  {author} {\bibinfo {author} {\bibfnamefont {S.}~\bibnamefont
  {Jin}}, \bibinfo {author} {\bibfnamefont {T.~H.}\ \bibnamefont {Tiefel}},
  \bibinfo {author} {\bibfnamefont {M.}~\bibnamefont {McCormack}}, \bibinfo
  {author} {\bibfnamefont {R.~A.}\ \bibnamefont {Fastnacht}}, \bibinfo {author}
  {\bibfnamefont {R.}~\bibnamefont {Ramesh}}, \ and\ \bibinfo {author}
  {\bibfnamefont {L.~H.}\ \bibnamefont {Chen}},\ }\href@noop {} {\bibfield
  {journal} {\bibinfo  {journal} {Science}\ }\textbf {\bibinfo {volume}
  {264}},\ \bibinfo {pages} {413} (\bibinfo {year} {1994})}\BibitemShut
  {NoStop}%
\bibitem [{\citenamefont {Uehara}\ \emph {et~al.}(1999)\citenamefont {Uehara},
  \citenamefont {Mori}, \citenamefont {Chen},\ and\ \citenamefont
  {Cheong}}]{ueh99}%
  \BibitemOpen
  \bibfield  {author} {\bibinfo {author} {\bibfnamefont {M.}~\bibnamefont
  {Uehara}}, \bibinfo {author} {\bibfnamefont {S.}~\bibnamefont {Mori}},
  \bibinfo {author} {\bibfnamefont {C.~H.}\ \bibnamefont {Chen}}, \ and\
  \bibinfo {author} {\bibfnamefont {S.~W.}\ \bibnamefont {Cheong}},\
  }\href@noop {} {\bibfield  {journal} {\bibinfo  {journal} {Nature}\ }\textbf
  {\bibinfo {volume} {399}},\ \bibinfo {pages} {560} (\bibinfo {year}
  {1999})}\BibitemShut {NoStop}%
\bibitem [{\citenamefont {Ahn}, \citenamefont {Lookman},\ and\ \citenamefont
  {Bishop}(2004)}]{ahn04}%
  \BibitemOpen
  \bibfield  {author} {\bibinfo {author} {\bibfnamefont {K.~H.}\ \bibnamefont
  {Ahn}}, \bibinfo {author} {\bibfnamefont {T.}~\bibnamefont {Lookman}}, \ and\
  \bibinfo {author} {\bibfnamefont {A.~R.}\ \bibnamefont {Bishop}},\
  }\href@noop {} {\bibfield  {journal} {\bibinfo  {journal} {Nature}\ }\textbf
  {\bibinfo {volume} {428}},\ \bibinfo {pages} {401} (\bibinfo {year}
  {2004})}\BibitemShut {NoStop}%
\bibitem [{\citenamefont {Biswas}\ \emph {et~al.}(2000)\citenamefont {Biswas},
  \citenamefont {Rajeswari}, \citenamefont {Srivastava}, \citenamefont {Li},
  \citenamefont {Venkatesan}, \citenamefont {Greene},\ and\ \citenamefont
  {Millis}}]{Bis00}%
  \BibitemOpen
  \bibfield  {author} {\bibinfo {author} {\bibfnamefont {A.}~\bibnamefont
  {Biswas}}, \bibinfo {author} {\bibfnamefont {M.}~\bibnamefont {Rajeswari}},
  \bibinfo {author} {\bibfnamefont {R.~C.}\ \bibnamefont {Srivastava}},
  \bibinfo {author} {\bibfnamefont {Y.~H.}\ \bibnamefont {Li}}, \bibinfo
  {author} {\bibfnamefont {T.}~\bibnamefont {Venkatesan}}, \bibinfo {author}
  {\bibfnamefont {R.~L.}\ \bibnamefont {Greene}}, \ and\ \bibinfo {author}
  {\bibfnamefont {A.~J.}\ \bibnamefont {Millis}},\ }\href@noop {} {\bibfield
  {journal} {\bibinfo  {journal} {Phys. Rev. B}\ }\textbf {\bibinfo {volume}
  {61}},\ \bibinfo {pages} {9665} (\bibinfo {year} {2000})}\BibitemShut
  {NoStop}%
\bibitem [{\citenamefont {Suzuki}\ \emph {et~al.}(1998)\citenamefont {Suzuki},
  \citenamefont {Hwang}, \citenamefont {Cheong}, \citenamefont {Siegrist},
  \citenamefont {van Dover}, \citenamefont {Asamitsu},\ and\ \citenamefont
  {Tokura}}]{Suz98}%
  \BibitemOpen
  \bibfield  {author} {\bibinfo {author} {\bibfnamefont {Y.}~\bibnamefont
  {Suzuki}}, \bibinfo {author} {\bibfnamefont {H.~Y.}\ \bibnamefont {Hwang}},
  \bibinfo {author} {\bibfnamefont {S.-W.}\ \bibnamefont {Cheong}}, \bibinfo
  {author} {\bibfnamefont {T.}~\bibnamefont {Siegrist}}, \bibinfo {author}
  {\bibfnamefont {R.~B.}\ \bibnamefont {van Dover}}, \bibinfo {author}
  {\bibfnamefont {A.}~\bibnamefont {Asamitsu}}, \ and\ \bibinfo {author}
  {\bibfnamefont {Y.}~\bibnamefont {Tokura}},\ } {\bibfield  {journal} {\bibinfo  {journal} {J. Appl.
  Phys.}\ }\textbf {\bibinfo {volume} {83}},\ \bibinfo {pages} {7064} (\bibinfo
  {year} {1998})}\BibitemShut {NoStop}%
\bibitem [{\citenamefont {Ward}\ \emph {et~al.}(2009)\citenamefont {Ward},
  \citenamefont {Budai}, \citenamefont {Gai}, \citenamefont {Tischler},
  \citenamefont {Yin},\ and\ \citenamefont {Shen}}]{War09}%
  \BibitemOpen
  \bibfield  {author} {\bibinfo {author} {\bibfnamefont {T.~Z.}\ \bibnamefont
  {Ward}}, \bibinfo {author} {\bibfnamefont {J.~D.}\ \bibnamefont {Budai}},
  \bibinfo {author} {\bibfnamefont {Z.}~\bibnamefont {Gai}}, \bibinfo {author}
  {\bibfnamefont {J.~Z.}\ \bibnamefont {Tischler}}, \bibinfo {author}
  {\bibfnamefont {L.}~\bibnamefont {Yin}}, \ and\ \bibinfo {author}
  {\bibfnamefont {J.}~\bibnamefont {Shen}},\ }\href@noop {} {\bibfield
  {journal} {\bibinfo  {journal} {Nat. Phys.}\ }\textbf {\bibinfo {volume}
  {5}},\ \bibinfo {pages} {885} (\bibinfo {year} {2009})}\BibitemShut {NoStop}%
\bibitem [{\citenamefont {Choi}\ \emph {et~al.}(2009)\citenamefont {Choi},
  \citenamefont {Marton}, \citenamefont {Jang}, \citenamefont {Moon},
  \citenamefont {Jeon}, \citenamefont {Shin}, \citenamefont {Seo},
  \citenamefont {Noh}, \citenamefont {Myung-Whun}, \citenamefont {Lee},\ and\
  \citenamefont {Lee}}]{Cho09}%
  \BibitemOpen
  \bibfield  {author} {\bibinfo {author} {\bibfnamefont {W.~S.}\ \bibnamefont
  {Choi}}, \bibinfo {author} {\bibfnamefont {Z.}~\bibnamefont {Marton}},
  \bibinfo {author} {\bibfnamefont {S.~Y.}\ \bibnamefont {Jang}}, \bibinfo
  {author} {\bibfnamefont {S.~J.}\ \bibnamefont {Moon}}, \bibinfo {author}
  {\bibfnamefont {B.~C.}\ \bibnamefont {Jeon}}, \bibinfo {author}
  {\bibfnamefont {J.~H.}\ \bibnamefont {Shin}}, \bibinfo {author}
  {\bibfnamefont {S.~S.~A.}\ \bibnamefont {Seo}}, \bibinfo {author}
  {\bibfnamefont {T.~W.}\ \bibnamefont {Noh}}, \bibinfo {author} {\bibfnamefont
  {K.}~\bibnamefont {Myung-Whun}}, \bibinfo {author} {\bibfnamefont {H.~N.}\
  \bibnamefont {Lee}}, \ and\ \bibinfo {author} {\bibfnamefont {Y.~S.}\
  \bibnamefont {Lee}},\ } {\bibfield  {journal}
  {\bibinfo  {journal} {J. Phys. D: Appl. Phys.}\ }\textbf {\bibinfo {volume}
  {42}},\ \bibinfo {pages} {165401} (\bibinfo {year} {2009})}\BibitemShut
  {NoStop}%
\bibitem [{\citenamefont {Prellier}, \citenamefont {Lecoeur},\ and\
  \citenamefont {Mercey}(2001)}]{Pre01}%
  \BibitemOpen
  \bibfield  {author} {\bibinfo {author} {\bibfnamefont {W.}~\bibnamefont
  {Prellier}}, \bibinfo {author} {\bibfnamefont {P.}~\bibnamefont {Lecoeur}}, \
  and\ \bibinfo {author} {\bibfnamefont {B.}~\bibnamefont {Mercey}},\ } {\bibfield  {journal}
  {\bibinfo  {journal} {J. Phys.: Condens. Matter}\ }\textbf {\bibinfo {volume}
  {13}},\ \bibinfo {pages} {R915} (\bibinfo {year} {2001})}\BibitemShut
  {NoStop}%
\bibitem [{\citenamefont {Marton}\ \emph {et~al.}(2010)\citenamefont {Marton},
  \citenamefont {Seo}, \citenamefont {Egami},\ and\ \citenamefont
  {Lee}}]{mar10}%
  \BibitemOpen
  \bibfield  {author} {\bibinfo {author} {\bibfnamefont {Z.}~\bibnamefont
  {Marton}}, \bibinfo {author} {\bibfnamefont {S.~S.~A.}\ \bibnamefont {Seo}},
  \bibinfo {author} {\bibfnamefont {T.}~\bibnamefont {Egami}}, \ and\ \bibinfo
  {author} {\bibfnamefont {H.~N.}\ \bibnamefont {Lee}},\ } {\bibfield  {journal} {\bibinfo  {journal}
  {J. Cryst. Growth}\ }\textbf {\bibinfo {volume} {312}},\ \bibinfo {pages}
  {2923 } (\bibinfo {year} {2010})}\BibitemShut {NoStop}%
\bibitem [{\citenamefont {Boschker}\ \emph {et~al.}(2009)\citenamefont
  {Boschker}, \citenamefont {Mathews}, \citenamefont {Houwman}, \citenamefont
  {Nishikawa}, \citenamefont {Vailionis}, \citenamefont {Koster}, \citenamefont
  {Rijnders},\ and\ \citenamefont {Blank}}]{Bos09}%
  \BibitemOpen
  \bibfield  {author} {\bibinfo {author} {\bibfnamefont {H.}~\bibnamefont
  {Boschker}}, \bibinfo {author} {\bibfnamefont {M.}~\bibnamefont {Mathews}},
  \bibinfo {author} {\bibfnamefont {E.~P.}\ \bibnamefont {Houwman}}, \bibinfo
  {author} {\bibfnamefont {H.}~\bibnamefont {Nishikawa}}, \bibinfo {author}
  {\bibfnamefont {A.}~\bibnamefont {Vailionis}}, \bibinfo {author}
  {\bibfnamefont {G.}~\bibnamefont {Koster}}, \bibinfo {author} {\bibfnamefont
  {G.}~\bibnamefont {Rijnders}}, \ and\ \bibinfo {author} {\bibfnamefont
  {D.~H.~A.}\ \bibnamefont {Blank}},\ } {\bibfield  {journal} {\bibinfo  {journal} {Phys.
  Rev. B}\ }\textbf {\bibinfo {volume} {79}},\ \bibinfo {pages} {214425}
  (\bibinfo {year} {2009})}\BibitemShut {NoStop}%
\bibitem [{\citenamefont {Bhattacharya}\ \emph {et~al.}(2008)\citenamefont
  {Bhattacharya}, \citenamefont {May}, \citenamefont {te~Velthuis},
  \citenamefont {Warusawithana}, \citenamefont {Zhai}, \citenamefont {Jiang},
  \citenamefont {Zuo}, \citenamefont {Fitzsimmons}, \citenamefont {Bader},\
  and\ \citenamefont {Eckstein}}]{Bha08}%
  \BibitemOpen
  \bibfield  {author} {\bibinfo {author} {\bibfnamefont {A.}~\bibnamefont
  {Bhattacharya}}, \bibinfo {author} {\bibfnamefont {S.~J.}\ \bibnamefont
  {May}}, \bibinfo {author} {\bibfnamefont {S.~G.~E.}\ \bibnamefont
  {te~Velthuis}}, \bibinfo {author} {\bibfnamefont {M.}~\bibnamefont
  {Warusawithana}}, \bibinfo {author} {\bibfnamefont {X.}~\bibnamefont {Zhai}},
  \bibinfo {author} {\bibfnamefont {B.}~\bibnamefont {Jiang}}, \bibinfo
  {author} {\bibfnamefont {J.-M.}\ \bibnamefont {Zuo}}, \bibinfo {author}
  {\bibfnamefont {M.~R.}\ \bibnamefont {Fitzsimmons}}, \bibinfo {author}
  {\bibfnamefont {S.~D.}\ \bibnamefont {Bader}}, \ and\ \bibinfo {author}
  {\bibfnamefont {J.~N.}\ \bibnamefont {Eckstein}},\ } {\bibfield  {journal} {\bibinfo  {journal}
  {Phys. Rev. Lett.}\ }\textbf {\bibinfo {volume} {100}},\ \bibinfo {pages}
  {257203} (\bibinfo {year} {2008})}\BibitemShut {NoStop}%
\bibitem [{\citenamefont {Smadici}\ \emph {et~al.}(2007)\citenamefont
  {Smadici}, \citenamefont {Abbamonte}, \citenamefont {Bhattacharya},
  \citenamefont {Zhai}, \citenamefont {Jiang}, \citenamefont {Rusydi},
  \citenamefont {Eckstein}, \citenamefont {Bader},\ and\ \citenamefont
  {Zuo}}]{Sma07}%
  \BibitemOpen
  \bibfield  {author} {\bibinfo {author} {\bibfnamefont {S.}~\bibnamefont
  {Smadici}}, \bibinfo {author} {\bibfnamefont {P.}~\bibnamefont {Abbamonte}},
  \bibinfo {author} {\bibfnamefont {A.}~\bibnamefont {Bhattacharya}}, \bibinfo
  {author} {\bibfnamefont {X.}~\bibnamefont {Zhai}}, \bibinfo {author}
  {\bibfnamefont {B.}~\bibnamefont {Jiang}}, \bibinfo {author} {\bibfnamefont
  {A.}~\bibnamefont {Rusydi}}, \bibinfo {author} {\bibfnamefont {J.~N.}\
  \bibnamefont {Eckstein}}, \bibinfo {author} {\bibfnamefont {S.~D.}\
  \bibnamefont {Bader}}, \ and\ \bibinfo {author} {\bibfnamefont {J.-M.}\
  \bibnamefont {Zuo}},\ }
  {\bibfield  {journal} {\bibinfo  {journal} {Phys. Rev. Lett.}\ }\textbf
  {\bibinfo {volume} {99}},\ \bibinfo {pages} {196404} (\bibinfo {year}
  {2007})}\BibitemShut {NoStop}%
\bibitem [{\citenamefont {Ohtomo}\ and\ \citenamefont {Hwang}(2004)}]{Oht04}%
  \BibitemOpen
  \bibfield  {author} {\bibinfo {author} {\bibfnamefont {A.}~\bibnamefont
  {Ohtomo}}\ and\ \bibinfo {author} {\bibfnamefont {H.~Y.}\ \bibnamefont
  {Hwang}},\ } {\bibfield  {journal}
  {\bibinfo  {journal} {Nature}\ }\textbf {\bibinfo {volume} {427}},\ \bibinfo
  {pages} {423} (\bibinfo {year} {2004})}\BibitemShut {NoStop}%
\bibitem [{\citenamefont {Ariando}\ \emph {et~al.}(2011)\citenamefont
  {Ariando}, \citenamefont {Wang}, \citenamefont {Baskaran}, \citenamefont
  {Liu}, \citenamefont {Huijben}, \citenamefont {Yi}, \citenamefont {Annadi},
  \citenamefont {Barman}, \citenamefont {Rusydi}, \citenamefont {Dhar},
  \citenamefont {Feng}, \citenamefont {Ding}, \citenamefont {Hilgenkamp},\ and\
  \citenamefont {Venkatesan}}]{Ari11}%
  \BibitemOpen
  \bibfield  {author} {\bibinfo {author} {\bibnamefont {Ariando}}, \bibinfo
  {author} {\bibfnamefont {X.}~\bibnamefont {Wang}}, \bibinfo {author}
  {\bibfnamefont {G.}~\bibnamefont {Baskaran}}, \bibinfo {author}
  {\bibfnamefont {Z.~Q.}\ \bibnamefont {Liu}}, \bibinfo {author} {\bibfnamefont
  {J.}~\bibnamefont {Huijben}}, \bibinfo {author} {\bibfnamefont {J.~B.}\
  \bibnamefont {Yi}}, \bibinfo {author} {\bibfnamefont {A.}~\bibnamefont
  {Annadi}}, \bibinfo {author} {\bibfnamefont {A.~R.}\ \bibnamefont {Barman}},
  \bibinfo {author} {\bibfnamefont {A.}~\bibnamefont {Rusydi}}, \bibinfo
  {author} {\bibfnamefont {S.}~\bibnamefont {Dhar}}, \bibinfo {author}
  {\bibfnamefont {Y.~P.}\ \bibnamefont {Feng}}, \bibinfo {author}
  {\bibfnamefont {J.}~\bibnamefont {Ding}}, \bibinfo {author} {\bibfnamefont
  {H.}~\bibnamefont {Hilgenkamp}}, \ and\ \bibinfo {author} {\bibfnamefont
  {T.}~\bibnamefont {Venkatesan}},\ } {\bibfield  {journal} {\bibinfo
  {journal} {Nat Commun}\ }\textbf {\bibinfo {volume} {2}},\ \bibinfo {pages}
  {188} (\bibinfo {year} {2011})}\BibitemShut {NoStop}%
\bibitem [{\citenamefont {Ward}\ \emph {et~al.}(2008)\citenamefont {Ward},
  \citenamefont {Liang}, \citenamefont {Fuchigami}, \citenamefont {Yin},
  \citenamefont {Dagotto}, \citenamefont {Plummer},\ and\ \citenamefont
  {Shen}}]{War08}%
  \BibitemOpen
  \bibfield  {author} {\bibinfo {author} {\bibfnamefont {T.~Z.}\ \bibnamefont
  {Ward}}, \bibinfo {author} {\bibfnamefont {S.}~\bibnamefont {Liang}},
  \bibinfo {author} {\bibfnamefont {K.}~\bibnamefont {Fuchigami}}, \bibinfo
  {author} {\bibfnamefont {L.~F.}\ \bibnamefont {Yin}}, \bibinfo {author}
  {\bibfnamefont {E.}~\bibnamefont {Dagotto}}, \bibinfo {author} {\bibfnamefont
  {E.~W.}\ \bibnamefont {Plummer}}, \ and\ \bibinfo {author} {\bibfnamefont
  {J.}~\bibnamefont {Shen}},\ }
  {\bibfield  {journal} {\bibinfo  {journal} {Phys. Rev. Lett.}\ }\textbf
  {\bibinfo {volume} {100}},\ \bibinfo {pages} {247204} (\bibinfo {year}
  {2008})}\BibitemShut {NoStop}%
\bibitem [{\citenamefont {Jeen}\ and\ \citenamefont {Biswas}(2011)}]{jee11}%
  \BibitemOpen
  \bibfield  {author} {\bibinfo {author} {\bibfnamefont {H.}~\bibnamefont
  {Jeen}}\ and\ \bibinfo {author} {\bibfnamefont {A.}~\bibnamefont {Biswas}},\
  } {\bibfield  {journal} {\bibinfo
  {journal} {Phys. Rev. B}\ }\textbf {\bibinfo {volume} {83}},\ \bibinfo
  {pages} {064408} (\bibinfo {year} {2011})}\BibitemShut {NoStop}%
\bibitem [{\citenamefont {Ju}\ \emph {et~al.}(1995)\citenamefont {Ju},
  \citenamefont {Gopalakrishnan}, \citenamefont {Peng}, \citenamefont {Li},
  \citenamefont {Xiong}, \citenamefont {Venkatesan},\ and\ \citenamefont
  {Greene}}]{ju95}%
  \BibitemOpen
  \bibfield  {author} {\bibinfo {author} {\bibfnamefont {H.~L.}\ \bibnamefont
  {Ju}}, \bibinfo {author} {\bibfnamefont {J.}~\bibnamefont {Gopalakrishnan}},
  \bibinfo {author} {\bibfnamefont {J.~L.}\ \bibnamefont {Peng}}, \bibinfo
  {author} {\bibfnamefont {Q.}~\bibnamefont {Li}}, \bibinfo {author}
  {\bibfnamefont {G.~C.}\ \bibnamefont {Xiong}}, \bibinfo {author}
  {\bibfnamefont {T.}~\bibnamefont {Venkatesan}}, \ and\ \bibinfo {author}
  {\bibfnamefont {R.~L.}\ \bibnamefont {Greene}},\ } {\bibfield  {journal} {\bibinfo  {journal} {Phys.
  Rev. B}\ }\textbf {\bibinfo {volume} {51}},\ \bibinfo {pages} {6143}
  (\bibinfo {year} {1995})}\BibitemShut {NoStop}%
\bibitem [{\citenamefont {Prellier}\ \emph {et~al.}(1999)\citenamefont
  {Prellier}, \citenamefont {Rajeswari}, \citenamefont {Venkatesan},\ and\
  \citenamefont {Greene}}]{Pre99}%
  \BibitemOpen
  \bibfield  {author} {\bibinfo {author} {\bibfnamefont {W.}~\bibnamefont
  {Prellier}}, \bibinfo {author} {\bibfnamefont {M.}~\bibnamefont {Rajeswari}},
  \bibinfo {author} {\bibfnamefont {T.}~\bibnamefont {Venkatesan}}, \ and\
  \bibinfo {author} {\bibfnamefont {R.~L.}\ \bibnamefont {Greene}},\ } {\bibfield  {journal} {\bibinfo  {journal} {Appl.
  Phys. Lett.}\ }\textbf {\bibinfo {volume} {75}},\ \bibinfo {pages} {1446}
  (\bibinfo {year} {1999})}\BibitemShut {NoStop}%
\bibitem [{\citenamefont {Leca}(2003)}]{lec03}%
  \BibitemOpen
  \bibfield  {author} {\bibinfo {author} {\bibfnamefont {V.}~\bibnamefont
  {Leca}},\ }\href@noop {} {Ph.D. thesis},\ \bibinfo  {school} {University of
  Twente} (\bibinfo {year} {2003})\BibitemShut {NoStop}%
\bibitem [{kei(2007)}]{kei07}%
  \BibitemOpen
  \href@noop {} {\emph {\bibinfo {title} {Nanotechnology measurement
  handbook}}}\ (\bibinfo  {publisher} {Keithley},\ \bibinfo {address}
  {Cleveland, OH},\ \bibinfo {year} {2007})\BibitemShut {NoStop}%
\bibitem [{\citenamefont {Dhakal}, \citenamefont {Tosado},\ and\ \citenamefont
  {Biswas}(2007)}]{dha07}%
  \BibitemOpen
  \bibfield  {author} {\bibinfo {author} {\bibfnamefont {T.}~\bibnamefont
  {Dhakal}}, \bibinfo {author} {\bibfnamefont {J.}~\bibnamefont {Tosado}}, \
  and\ \bibinfo {author} {\bibfnamefont {A.}~\bibnamefont {Biswas}},\
  }\href@noop {} {\bibfield  {journal} {\bibinfo  {journal} {Phys. Rev. B}\
  }\textbf {\bibinfo {volume} {75}},\ \bibinfo {pages} {092404} (\bibinfo
  {year} {2007})}\BibitemShut {NoStop}%
\bibitem [{\citenamefont {Collado}\ \emph {et~al.}(2003)\citenamefont
  {Collado}, \citenamefont {Frontera}, \citenamefont {García-Muñoz},
  \citenamefont {Ritter}, \citenamefont {Brunelli},\ and\ \citenamefont
  {Aranda}}]{Col03}%
  \BibitemOpen
  \bibfield  {author} {\bibinfo {author} {\bibfnamefont {J.~A.}\ \bibnamefont
  {Collado}}, \bibinfo {author} {\bibfnamefont {C.}~\bibnamefont {Frontera}},
  \bibinfo {author} {\bibfnamefont {J.~L.}\ \bibnamefont {García-Muñoz}},
  \bibinfo {author} {\bibfnamefont {C.}~\bibnamefont {Ritter}}, \bibinfo
  {author} {\bibfnamefont {M.}~\bibnamefont {Brunelli}}, \ and\ \bibinfo
  {author} {\bibfnamefont {M.~A.~G.}\ \bibnamefont {Aranda}},\ } {\bibfield  {journal} {\bibinfo  {journal} {Chem. Mater.}\
  }\textbf {\bibinfo {volume} {15}},\ \bibinfo {pages} {167} (\bibinfo {year}
  {2003})}\BibitemShut {NoStop}%
\bibitem [{\citenamefont {Tosado}, \citenamefont {Dhakal},\ and\ \citenamefont
  {Biswas}(2009)}]{tos09}%
  \BibitemOpen
  \bibfield  {author} {\bibinfo {author} {\bibfnamefont {J.}~\bibnamefont
  {Tosado}}, \bibinfo {author} {\bibfnamefont {T.}~\bibnamefont {Dhakal}}, \
  and\ \bibinfo {author} {\bibfnamefont {A.}~\bibnamefont {Biswas}},\
  }\href@noop {} {\bibfield  {journal} {\bibinfo  {journal} {J. Phys.: Condens.
  Mat.}\ }\textbf {\bibinfo {volume} {21}},\ \bibinfo {pages} {192203}
  (\bibinfo {year} {2009})}\BibitemShut {NoStop}%
\bibitem [{\citenamefont {Christen}\ and\ \citenamefont {Eres}(2008)}]{Chr08}%
  \BibitemOpen
  \bibfield  {author} {\bibinfo {author} {\bibfnamefont {H.~M.}\ \bibnamefont
  {Christen}}\ and\ \bibinfo {author} {\bibfnamefont {G.}~\bibnamefont
  {Eres}},\ }
  {\bibfield  {journal} {\bibinfo  {journal} {J. Phys.: Condens. Matter}\
  }\textbf {\bibinfo {volume} {20}},\ \bibinfo {pages} {264005} (\bibinfo
  {year} {2008})}\BibitemShut {NoStop}%
\bibitem [{\citenamefont {Urushibara}\ \emph {et~al.}(1995)\citenamefont
  {Urushibara}, \citenamefont {Moritomo}, \citenamefont {Arima}, \citenamefont
  {Asamitsu}, \citenamefont {Kido},\ and\ \citenamefont {Tokura}}]{Uru95}%
  \BibitemOpen
  \bibfield  {author} {\bibinfo {author} {\bibfnamefont {A.}~\bibnamefont
  {Urushibara}}, \bibinfo {author} {\bibfnamefont {Y.}~\bibnamefont
  {Moritomo}}, \bibinfo {author} {\bibfnamefont {T.}~\bibnamefont {Arima}},
  \bibinfo {author} {\bibfnamefont {A.}~\bibnamefont {Asamitsu}}, \bibinfo
  {author} {\bibfnamefont {G.}~\bibnamefont {Kido}}, \ and\ \bibinfo {author}
  {\bibfnamefont {Y.}~\bibnamefont {Tokura}},\ } {\bibfield  {journal} {\bibinfo  {journal} {Phys.
  Rev. B}\ }\textbf {\bibinfo {volume} {51}},\ \bibinfo {pages} {14103}
  (\bibinfo {year} {1995})}\BibitemShut {NoStop}%
\bibitem [{\citenamefont {Schiffer}\ \emph {et~al.}(1995)\citenamefont
  {Schiffer}, \citenamefont {Ramirez}, \citenamefont {Bao},\ and\ \citenamefont
  {Cheong}}]{Sch95}%
  \BibitemOpen
  \bibfield  {author} {\bibinfo {author} {\bibfnamefont {P.}~\bibnamefont
  {Schiffer}}, \bibinfo {author} {\bibfnamefont {A.~P.}\ \bibnamefont
  {Ramirez}}, \bibinfo {author} {\bibfnamefont {W.}~\bibnamefont {Bao}}, \ and\
  \bibinfo {author} {\bibfnamefont {S.-W.}\ \bibnamefont {Cheong}},\ } {\bibfield  {journal} {\bibinfo
  {journal} {Phys. Rev. Lett.}\ }\textbf {\bibinfo {volume} {75}},\ \bibinfo
  {pages} {3336} (\bibinfo {year} {1995})}\BibitemShut {NoStop}%
\bibitem [{\citenamefont {Orgiani}\ \emph {et~al.}(2010)\citenamefont
  {Orgiani}, \citenamefont {Ciancio}, \citenamefont {Galdi}, \citenamefont
  {Amoruso},\ and\ \citenamefont {Maritato}}]{Org10}%
  \BibitemOpen
  \bibfield  {author} {\bibinfo {author} {\bibfnamefont {P.}~\bibnamefont
  {Orgiani}}, \bibinfo {author} {\bibfnamefont {R.}~\bibnamefont {Ciancio}},
  \bibinfo {author} {\bibfnamefont {A.}~\bibnamefont {Galdi}}, \bibinfo
  {author} {\bibfnamefont {S.}~\bibnamefont {Amoruso}}, \ and\ \bibinfo
  {author} {\bibfnamefont {L.}~\bibnamefont {Maritato}},\ } {\bibfield  {journal} {\bibinfo  {journal} {Appl. Phys.
  Lett.}\ }\textbf {\bibinfo {volume} {96}},\ \bibinfo {eid} {032501} (\bibinfo
  {year} {2010})}\BibitemShut {NoStop}%
\bibitem [{\citenamefont {Hwang}\ \emph {et~al.}(1996)\citenamefont {Hwang},
  \citenamefont {Cheong}, \citenamefont {Ong},\ and\ \citenamefont
  {Batlogg}}]{Hwa96}%
  \BibitemOpen
  \bibfield  {author} {\bibinfo {author} {\bibfnamefont {H.~Y.}\ \bibnamefont
  {Hwang}}, \bibinfo {author} {\bibfnamefont {S.-W.}\ \bibnamefont {Cheong}},
  \bibinfo {author} {\bibfnamefont {N.~P.}\ \bibnamefont {Ong}}, \ and\
  \bibinfo {author} {\bibfnamefont {B.}~\bibnamefont {Batlogg}},\ } {\bibfield  {journal} {\bibinfo
  {journal} {Phys. Rev. Lett.}\ }\textbf {\bibinfo {volume} {77}},\ \bibinfo
  {pages} {2041} (\bibinfo {year} {1996})}\BibitemShut {NoStop}%
\bibitem [{\citenamefont {Taniyama}, \citenamefont {Yamasaki},\ and\
  \citenamefont {Yamazaki}(2002)}]{Tan02}%
  \BibitemOpen
  \bibfield  {author} {\bibinfo {author} {\bibfnamefont {T.}~\bibnamefont
  {Taniyama}}, \bibinfo {author} {\bibfnamefont {M.}~\bibnamefont {Yamasaki}},
  \ and\ \bibinfo {author} {\bibfnamefont {Y.}~\bibnamefont {Yamazaki}},\
  } {\bibfield  {journal} {\bibinfo
  {journal} {Applied Physics Letters}\ }\textbf {\bibinfo {volume} {81}},\
  \bibinfo {pages} {4562} (\bibinfo {year} {2002})}\BibitemShut {NoStop}%
\bibitem [{\citenamefont {Salamon}\ and\ \citenamefont {Jaime}(2001)}]{Sal01}%
  \BibitemOpen
  \bibfield  {author} {\bibinfo {author} {\bibfnamefont {M.~B.}\ \bibnamefont
  {Salamon}}\ and\ \bibinfo {author} {\bibfnamefont {M.}~\bibnamefont
  {Jaime}},\ } {\bibfield  {journal}
  {\bibinfo  {journal} {Rev. Mod. Phys.}\ }\textbf {\bibinfo {volume} {73}},\
  \bibinfo {pages} {583} (\bibinfo {year} {2001})}\BibitemShut {NoStop}%
\bibitem [{\citenamefont {Ziese}\ and\ \citenamefont
  {Srinitiwarawong}(1998)}]{Zie98a}%
  \BibitemOpen
  \bibfield  {author} {\bibinfo {author} {\bibfnamefont {M.}~\bibnamefont
  {Ziese}}\ and\ \bibinfo {author} {\bibfnamefont {C.}~\bibnamefont
  {Srinitiwarawong}},\ } {\bibfield
  {journal} {\bibinfo  {journal} {Phys. Rev. B}\ }\textbf {\bibinfo {volume}
  {58}},\ \bibinfo {pages} {11519} (\bibinfo {year} {1998})}\BibitemShut
  {NoStop}%
\bibitem [{\citenamefont {Yun}(2008)}]{yun08}%
  \BibitemOpen
  \bibfield  {author} {\bibinfo {author} {\bibfnamefont {S.~H.}\ \bibnamefont
  {Yun}},\ }\href@noop {} {Ph.D. thesis},\ \bibinfo  {school} {University of
  Florida} (\bibinfo {year} {2008})\BibitemShut {NoStop}%
\end{thebibliography}
%

\end{document}